# KEY LESSONS LEARNED FROM WORKING DURING COVID-19 ON A PROJECT IN THE WORLD'S BIGGEST REFUGEE CAMP

Faheem Hussain, Arizona State University, USA, faheem.hussain@asu.edu

Suzana Brown, SUNY Korea, South Korea, suzana.brown@sunykorea.ac.kr

**Abstract:** Using a case study structure, this research-in-progress paper elaborates the struggles of working on a humanitarian project during the Covid-19 period. The authors identify six specific challenges and propose innovations to address each of these challenges. The challenges are the following: supply chain, design of solutions, human resource development, connectivity, and user data collection. This unprecedented situation has been a testing ground for new innovative solutions for work in conflict zones.

## 1.     INTRODUCTION

This "research-in-progress" paper is based on the findings from ongoing humanitarian technology research in the world's biggest refugee camp for Rohingyas, on the border between Myanmar and Bangladesh.

## 2.     BACKGROUND

Rohingyas are arguably the most persecuted people at present with the majority of them living as refugees in Bangladesh, after fleeing the systematic ethnic cleansing in their native country Myanmar. Out of the 1.3 million Rohingya refugees in Bangladesh, many are facing challenges with physical mobility. The main cause behind such disabilities among Rohingya refugees is little or no health services in their native villages in Myanmar and long-term injuries sustained while fleeing the conflict.

Rohingya camps are located in the second poorest district in the country, Cox's Bazar (Vince, 2019). The refugees live on a deforested hillside, densely packed, sleeping on mats in shacks made from plastic woven walls and roofed with plastic sheets. The refugee "city" has very few latrines and freshwater standpipes (Vince, 2020).

In addition, the standard request during Covid-19 to socially distance is nearly impossible in those camps (Raju, & Ayeb-Karlsson, 2020). Refugee camps and slums are constructed as temporary places with very little extra space. Refugees already make up a vulnerable segment of society and people residing in refugee camps have a higher risk to be affected by an outbreak of Covid-19 (Vonen et al., 2020).

In such an environment and during the current pandemic, a research group from SUNY Korea, Hellenic Mediterranean University, and Arizona State University are working in collaboration with a partner NGO of Bangladesh, Young Power in Social Action (YPSA) to address this accessibility issue. The team is currently developing and testing an alternative design for the rubber shoe of mobility aids such as crutches and canes. This cost-effective design would function better in challenging environments such as unprepared and unpaved surfaces. The team estimates this new design will allow users to navigate with increased stability and thus result in an improved quality of life.

# 3. METHODOLOGY

Using a case study structure, this research-in-progress paper elaborates these aforementioned struggles and innovations to address each of these challenges. Beyond this specific project, this paper aims to contribute to the ongoing conversation on conducting international field research during the time of COVID-19. This research recommends a set of action items to overcome some of the major obstacles against humanitarian technology field research in #NewNormal.

## 3.1. Supply chain

This project was based on the use of 3D printers which are novel for the Bangladeshi location. In relatively normal times this issue would have been overcome with an import from either China or India. During Covid-19 not only travel but also distribution of goods has been disrupted with essential supplies having priority. After several months of intense negotiation, one 3D printer was imported from China during a period when the government relaxed restrictions on imports.

## 3.2. Solution design

Engineers and designers were not able to go to the field as planned and furthermore, they were physically distributed between 3 countries with variable travel restrictions. Quarantine made it impossible for them to travel even if allowed by the local authorities. Even within the country travel was difficult which limited local partner movement to and from the camp areas.

## 3.3 Human resource development

The initial plan was to have face-to-face training which had to be redesigning to online training. Because of the different time zones many of the sessions had to be asynchronous.

## 3.4 Connectivity

Bangladesh has relatively poor connectivity but the refugee camps have zero connectivity. Research team members had to employ some novel approaches in communication that could later be used in some other conflict zones with zero connectivity.

## 3.5 User data collection

Data collection and user feedback had to be reinvented and low-end smartphones have been used to record videos from the field for researchers to be able to evaluate their product.

## 3.6 Unpredictable policy changes

Finally, constant policy changes in almost every country made it impossible to make longer-term plans. Those include regulations regarding access to the camp areas; who is allowed to visit the host communities; regulation regarding shipment during COVID-19, etc.

# 4. CONCLUSIONS

Research centered on a humanitarian project poses its own challenges. However, during a health pandemic, all the attenuating circumstances are becoming more complicated. In this short "research-in-progress" paper we inform readers about a number of issues arising while conducting international field research during the time of COVID-19 in the world's largest refugee camp in Cox's Bazar, Bangladesh. We group those issues as follows: supply chain, connectivity, human development, unpredictable policy changes, user data collection, and solution design. In our future work, we will present insights on if and how such issues have been resolved.

# REFERENCES AND CITATIONS